\documentclass[12pt]{iopart}
\usepackage{iopams}
\usepackage{epsfig}

\begin{document}
\title[Costly hide and seek pays: Unexpected consequences of deceit in a social dilemma]{Costly hide and seek pays: Unexpected consequences of deceit in a social dilemma}

\author{Attila Szolnoki$^1$ and Matja{\v z} Perc$^{2,3}$}
\address{$^1$Institute of Technical Physics and Materials Science, Research Centre for Natural Sciences, Hungarian Academy of Sciences, P.O. Box 49, H-1525 Budapest, Hungary\\
$^2$Department of Physics, Faculty of Natural Sciences and Mathematics, University of Maribor, Koro{\v s}ka  cesta 160, SI-2000 Maribor, Slovenia\\
$^3$Department of Physics, Faculty of Science, King Abdulaziz University, Jeddah, Saudi Arabia}
\ead{szolnoki.attila@ttk.mta.hu, matjaz.perc@uni-mb.si}

\begin{abstract}
Deliberate deceptiveness intended to gain an advantage is commonplace in human and animal societies. In a social dilemma, an individual may only pretend to be a cooperator to elicit cooperation from others, while in reality he is a defector. With this as motivation, we study a simple variant of the evolutionary prisoner's dilemma game entailing deceitful defectors and conditional cooperators that lifts the veil on the impact of such two-faced behavior. Defectors are able to hide their true intentions at a personal cost, while conditional cooperators are probabilistically successful at identifying defectors and act accordingly. By focusing on the evolutionary outcomes in structured populations, we observe a number of unexpected and counterintuitive phenomena. We show that deceitful behavior may fare better if it is costly, and that a higher success rate of identifying defectors does not necessarily favor cooperative behavior. These results are rooted in the spontaneous emergence of cycling dominance and spatial patterns that give rise to fascinating phase transitions, which in turn reveal the hidden complexity behind the evolution of deception.
\end{abstract}

\pacs{87.23.Ge, 89.75.Fb, 89.65.-s}
\maketitle

\section{Introduction}
\label{intro}
Natural selection favors the fittest under adversity and testing conditions. According to Darwin's \textit{The Origin of Species}, organisms therefore change gradually over time to give rise to the astonishing diversity of life that is on display today \cite{carroll_06}. Sometimes, the most effective change is pretending to be someone or something one is not. In the animal world, mimicry is common to provide evolutionary advantages through an increased ability to escape predation or by elevating chances of predatory success \cite{owen_80}. The mimics and the species they are trying to fool are in an arms race, each trying to optimize their chances of survival while having to accommodate additional costs. Beautiful examples of mimicry include the Pandora sphinx moth that looks like a dead leaf to avoid detection, the Flower Mantis that mimic flowers to lure prey, and the many insects that have adopted the yellow and black stripes common to bees and wasps to fool others they are precisely that. Cuckoos are particularly cunning and famous for their breeding behavior. A female cuckoo lays its egg in the nest of a completely different species of bird, simply because it wants to avoid spending energy on raising offspring. An important mechanism for getting away with such behavior lies in the ability of cuckoos to cleverly deceive their host \cite{davies_00}. First, the egg the cuckoo lays is very similar to the host species eggs, and second, when the cuckoo chick hatches, it mimics the calls made by a whole brood of the host species chicks. In human societies, the ways of deception are of course even more cunning and elaborate. Our advanced intellectual abilities convey to us an impressive array of different strategies and actions by means of which we may fool others into a different reality. Obviously, some forms of trickery involve little to no additional costs, while others impose a significant burden on the practitioners.

Here we study the impact of such deceitful behavior on the evolution of cooperation in a social dilemma. Like deceptiveness intended to gain an advantage, situations that constitute social dilemmas are common in human and animal societies. In general, a social dilemma implies that the collective wellbeing is at odds with individual success \cite{macy_pnas02}. Individuals are therefore tempted to defect and maximize their own profit, but at the same time neglecting negative consequences such behavior has for the society as a whole. A frequently quoted outcome of such selfishness is the ``tragedy of the commons'' \cite{hardin_g_s68}. Indeed, the evolution of cooperation remains an evolutionary riddle \cite{axelrod_84,nowak_11}, and it is one of the most important challenges to Darwin's theory of evolution and natural selection. If during the course of evolution only the fittest survive, why should one sacrifice individual fitness for the benefit of unrelated others? While there is no single answer to this question, several mechanisms are known that promote cooperative behavior \cite{nowak_s06}.

Evolutionary game theory \cite{weibull_95,hofbauer_98,nowak_06} is long established as the theory of choice for studying the evolution of cooperation among selfish individuals, and likely the most frequently studied social dilemma is the prisoner's dilemma game \cite{fudenberg_e86,nowak_n93,szabo_pr07,santos_prl05,imhof_pnas05,tanimoto_pre07,gomez-gardenes_prl07,poncela_njp07,fu_pre08b,lee_s_prl11,antonioni_pone11,fu_srep12,tanimoto_pre12,yang_njp14}. Defection is the Nash equilibrium of the game, as it is the optimal strategy regardless of the strategy of the opponent. Beyond the consideration of cooperators and defectors as the simplest competing strategies, one of the most recent developments is the introduction of more advanced strategies that engage in evolutionary social dilemmas \cite{szolnoki_epl11,brede_epl11,fu_prsb11,vukov_njp12,brede_acs12,szolnoki_pre12,fu_srep12b,sasaki_dga14,szolnoki_pre13,exadaktylos_srep13}. Typically, individuals are endowed with cognitive skills, for example enabling them to identify the actions of other players or to learn from the failures made in previous rounds of the game.
Along this line, unconditional strategies --- cooperators that always cooperate and defectors that always defect --- constitute a simplification that deserves further exploration since it is a fact that individuals, be it humans or animals, will likely behave differently under different circumstances \cite{szolnoki_pre12}.

This invites the introduction of conditional strategies and deceptiveness \cite{trivers_11, mcnally_prsb13}, both of which we accommodate in the presently studied variant of the evolutionary prisoner's dilemma game. In particular, we introduce conditional cooperators ($C$) that cooperate only with other cooperators but defect otherwise, and we introduce deceitful defectors ($X$) that only pretend to be cooperative. In this way, we focus only on the ``darker'' side of deception, although it is worth emphasizing that prosocial lies with positive motivation have also been studied \cite{iniguez_prsb14}. Consequently, we allow defectors to go beyond pure defection ($D$), thus potentially providing a competitive answer to conditional cooperators. In addition to the temptation to defect $r$, however, these modifications introduce two additional parameters. Namely the probability $p$ that a conditional cooperator will correctly identify a pure defector and avoid being exploited, and the cost $\gamma$ that deceitful defectors need to bear in order to successfully belie cooperation. For further details we refer to the Model section. The questions we seek to answer within this theoretical framework are: What conditions allow the evolution of deception? How large can affordable $\gamma$ values be? And what is the role of the effectiveness of conditional cooperators in identifying defectors? As we will show, the answers to these questions are far from trivial. While low detection probabilities help defectors and high hiding costs obviously work against the effectiveness of deception, much more unexpectedly, we will also show how deceitful behavior may fare better if it is costly, and how a higher success rate of identifying defectors does not necessarily favor cooperative behavior. These results are due to the spontaneous emergence of cyclic dominance and self-organized pattern formation, both of which give rise to continuous and discontinuous phase transitions that highlight the complexity behind the evolution of deception.

\section{Model}

\subsection{Deceitful defectors and conditional cooperators}
\noindent We consider a simple three-strategy social dilemma game where players can be deceitful defectors ($X$), conditional cooperators ($C$), or pure defectors ($D$). The payoffs among strategies are defined by the matrices

\vspace{0.3cm}\centerline{\begin{tabular}{r|c c c}
{\bf A} & $D$ & $C$ & $X$\\
\hline
$D$ & 0 & 0 & 0\\
$C$ & 0 & 1 & $S$\\
$X$ & \,\,$-\gamma$ \,\,\,\,\,& $(T-\gamma)$ \,\,\,\,\,& $-\gamma$\\
\end{tabular}
\hspace{0.6cm}
and
\hspace{0.6cm}
\begin{tabular}{r|c c c}
{\bf B} & $D$ & $C$ & $X$\\
\hline
$D$ & 0 & $T$ & 0\\
$C$ & $S$ & 1 & $S$\\
$X$ & \,\,$-\gamma$ \,\,\,\,\,& $(T-\gamma)$ \,\,\,\,\,& $-\gamma$\\
\end{tabular}\,.}\vspace{0.3cm}

\noindent We use the payoff matrix {\bf A} with probability $p$ and the payoff matrix {\bf B} with probability $1-p$. In matrix {\bf A} the conditional cooperator correctly identifies pure defectors and acts as a defector itself, while in  matrix {\bf B} the conditional cooperator fails to identify pure defectors and thus decides to cooperate. In the latter case, strategies $C$ and $D$ are simply unconditional cooperation and defection. Importantly, as a specific case of a more general model \cite{mcnally_prsb13}, conditional cooperators always cooperate with deceitful defectors, as the latter invest $\gamma$ specifically to that effect. If we would allow conditional cooperators to reveal also the deceptiveness of deceitful defectors the cost $\gamma$ would simply always constitute an evolutionary disadvantage.

Without loosing generality, we use the temptation to defect $T=1+r$ and the sucker's payoff $S=-r$, thus building upon the traditional prisoner's dilemma formulation of an evolutionary social dilemma game. Here the parameter $r>0$ determines the strength of the dilemma, and in what follows, we will present results for $r=0.3$ and $r=0.7$, being representative for a moderate and a strong prisoner's dilemma, respectively.

\subsection{Monte Carlo simulations}
\noindent We perform Monte Carlo simulations of the evolutionary social dilemma on a square lattice of size $L^2$ with periodic boundary conditions. The square lattice is the simplest of networks that allows us to go beyond well-mixed populations, and as such it enables us to take into account the fact that the interactions among competing species are often structured rather than random. By using the square lattice, we also continue a long-standing tradition that has begun with the work of Nowak and May \cite{nowak_n92b}, who were the first to show that the most striking differences in the outcome of an evolutionary game emerge when the assumption of a well-mixed population is abandoned for the usage of a structured population \cite{szabo_pr07,perc_bs10,santos_pnas06,santos_jtb12,fortunato_pr10,pinheiro_njp12,sun_njp10}.

Initially, each player on site $x$ is designated either as a deceitful defector ($s_x = X$), a conditional cooperator ($s_x = C$) or a pure defector ($s_x = D$) with equal probability. The Monte Carlo simulations comprise the following elementary steps. First, a randomly selected player $x$ acquires its payoff $\Pi_x$ by playing the game with its four nearest neighbors. Next, one randomly chosen neighbor, denoted by $y$, also acquires its payoff $\Pi_y$ in the same way. Lastly, player $y$ adopts the strategy of player $x$ with the probability
\begin{equation}
w(s_x \to s_y)=\frac{1}{1+\exp[(\Pi_{y}-\Pi_{x})/(K)]}\ \,,
\label{fermi}
\end{equation}
where $K$ determines the level of uncertainty in the Fermi function \cite{szabo_pr07}. The latter can be attributed to errors in judgment due to mistakes and external influences that affect the evaluation of the opponent. We use $K=0.1$ throughout this work, which implies that better performing players are readily imitated, but it is not impossible to adopt the strategy of a player performing worse. We note that the main results are robust to variations of $K$ and the strategy adoption rule, such as choosing the best or the better performing neighbor for the imitation, and they remain qualitatively valid also on other lattices and random networks where the degree distribution remains unchanged but the links are uncorrelated.

Each full Monte Carlo step (MCS) gives a chance to every player to change its strategy once on average. Depending on the proximity to phase transition points and the typical size of emerging spatial patterns, we have varied the linear size of the lattice from $L=400$ to $L=6000$ and the relaxation time from $10^3$ to $10^5$ MCS to obtain solutions that are valid in the large system size limit, and to ensure that the statistical error is comparable with the size of the symbols in the figures. Importantly, even at such a large system size ($L=6000$), for certain parameter values close to discontinuous phase transition points, the random initial state may not necessarily yield a relaxation towards the most stable solution of the game. To verify the stability of different subsystem solutions, we have therefore applied also prepared initial states (see for example Fig.~10 in \cite{szolnoki_pre11b}), and we have followed the same procedure as applied previously in \cite{szolnoki_prl12,szolnoki_prx13}.

\section{Results}

\subsection{Well-mixed populations}
Before presenting the main results obtained in structured populations, we briefly describe the evolutionary outcomes in well-mixed populations, where players interact with the whole population and choose competitors randomly \cite{mcnally_prsb13}.

In the $p=1$ limit, where pure defectors are always uncovered, strategy $C$ is superior to strategy $D$. This relation, however, is reversed if $p<\frac{r}{1+r}$. For $p$ values in-between, a bistable competition between $C$ and $D$ is possible, whereby the final outcome depends on the initial concentration of the competing strategies. In other words, $C$ and $D$ cannot coexist regardless of the value of $p$. The coexistence of $C$ and $X$ is also impossible, because deceitful defectors dominate cooperators if $\gamma<r$, while otherwise strategy $C$ is superior to strategy $X$. Lastly, we note that $D$ always beats $X$ because the latter have to bear the additional cost $\gamma$.

As a consequence, strategy $C$ prevails in the whole population if the values of $p$ and $\gamma$ are sufficiently high. Similarly, the full $D$ phase is attainable in the small $p$ limit. A mixed equilibrium, where all three competing strategies coexist, is also possible if $(1-p)r(1+r)<\gamma<p(1+r)$ and $\gamma<r$ are fulfilled simultaneously. Although these results already provide useful insight into the impact and evolutionary stability of deception, we next focus on studying evolutionary outcomes in structured populations.

\subsection{Structured populations}

\begin{figure}
\centerline{\epsfig{file=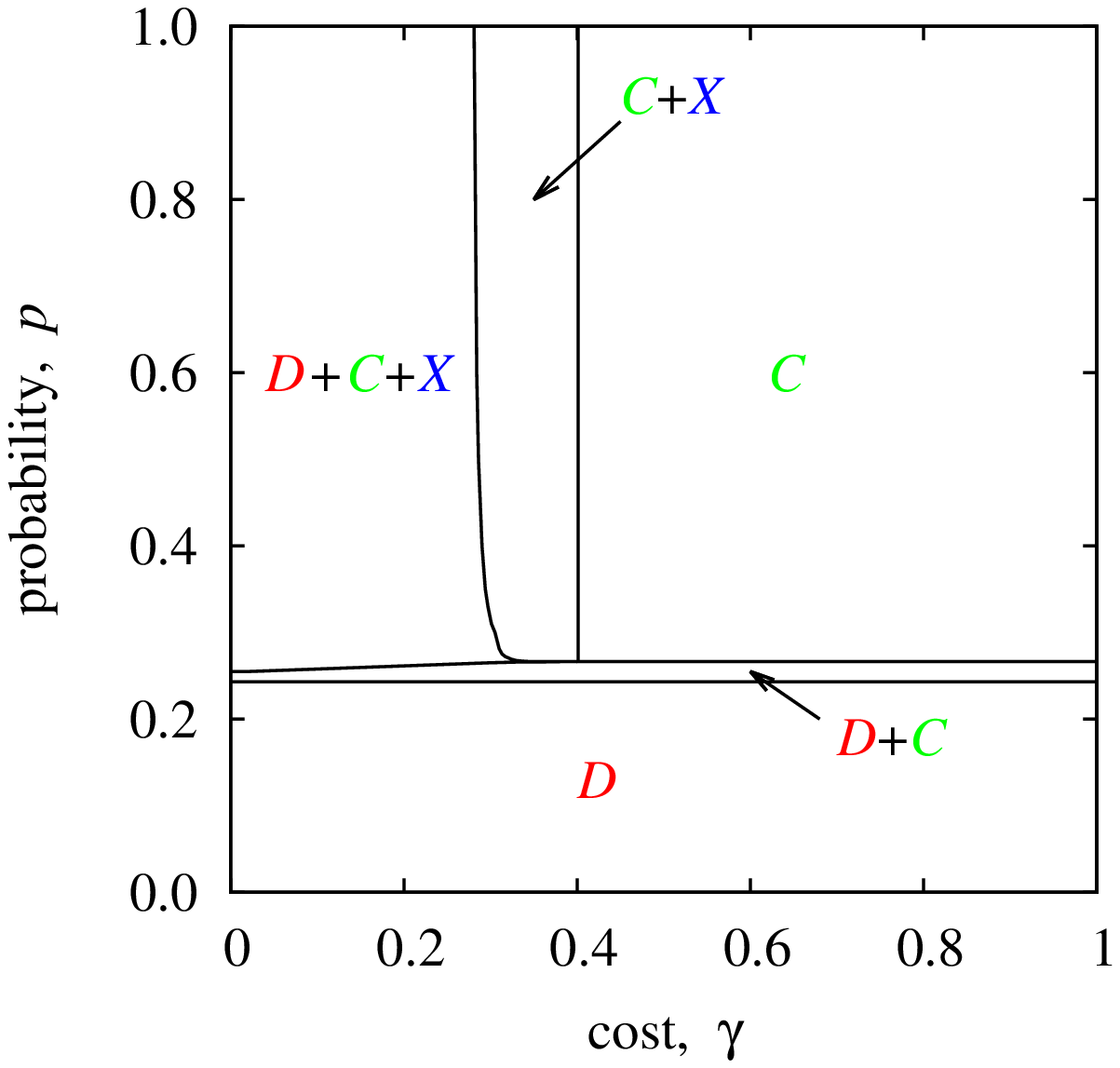,width=8cm}\epsfig{file=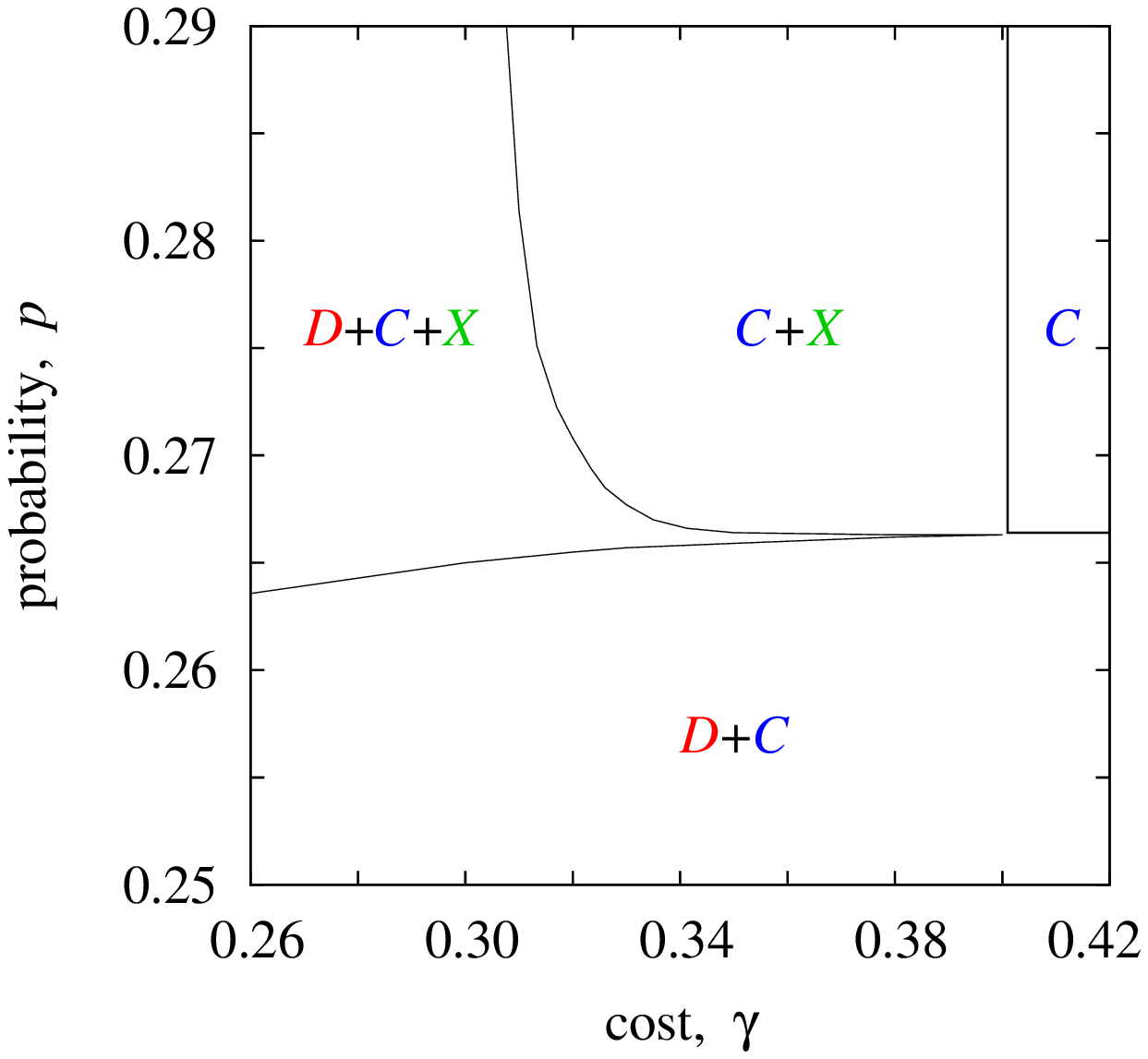,width=8cm}}
\caption{Full $\gamma-p$ phase diagram, as obtained for $r=0.3$. Solid lines denote continuous phase transitions. The vertical resolution hides the intricate structure of the phase diagram for intermediate values of $\gamma$ and $p$, which we therefore show enlarged in the right panel. Stable solutions include the three-strategy $C+D+X$ phase, two-strategy $C+D$ and $C+X$ phases, as well as the absorbing $D$ and $C$ phase.}
\label{phasea}
\end{figure}

In structured populations, we first focus on the moderate limit of the prisoner's dilemma game that is obtained for $r=0.3$. The left panel of Fig.~\ref{phasea} shows the full $\gamma-p$ phase diagram, which describes all possible stable solutions. Evidently, the richness of solutions is greater than in well-mixed populations. In general, small detection probabilities, when conditional cooperators frequently fail to correctly identify pure defectors, are beneficial for the evolution of defection, thus yielding an absorbing $D$ phase as the only stable solution in this region of the phase diagram, regardless of the cost of deception. As the effectiveness of conditional cooperators increases, the pure $D$ phase transforms into a two-strategy $D+C$ phase. This solution, which is absent in well-mixed populations, is due to the aggregation of cooperators into compact clusters, by means of which a stable coexistence of the two strategies becomes possible within a narrow band of $p$ (see also the right panel of Fig.~\ref{phasea}). This is a purely spatial effect that is rooted in network reciprocity \cite{nowak_n92b}. If both $\gamma$ and $p$ are large, the $D+C$ phase terminates into an absorbing $C$ phase, while for sufficiently low values of $\gamma$ deceitful defectors become viable, either through the emergence of a two-strategy $C+X$ phase or the emergence of a three-strategy $C+D+X$ phase. Within the latter the competing strategies may dominate each other cyclically, although the stable coexistence of all three strategies in the $D+C+X$ phase does not always involve cyclic dominance. Several aspects of these results are counterintuitive and unexpected. Foremost, one would expect that decreasing values of $p$ will impair the evolution of $C$, and that increasing values of $\gamma$ will be detrimental for $X$. But this is not necessarily the case. In fact, as the value of $p$ decreases, the first to die out are the deceitful defectors, giving way to a mixed $C+D$ phase. Moreover, as $\gamma$ increases, the first to vanish from the three-strategy phase are the pure defectors, thus yielding the $C+X$ phase. If $\gamma> 0.401$ and $p > 0.2664$, then only $C$ can survive. Interestingly, the two two-strategy phases are always separated by the three-strategy $D+C+X$ phase, as illustrated clearly in the enlarged part of the phase diagram depicted in the right panel of Fig.~\ref{phasea}.

\begin{figure}
\centerline{\epsfig{file=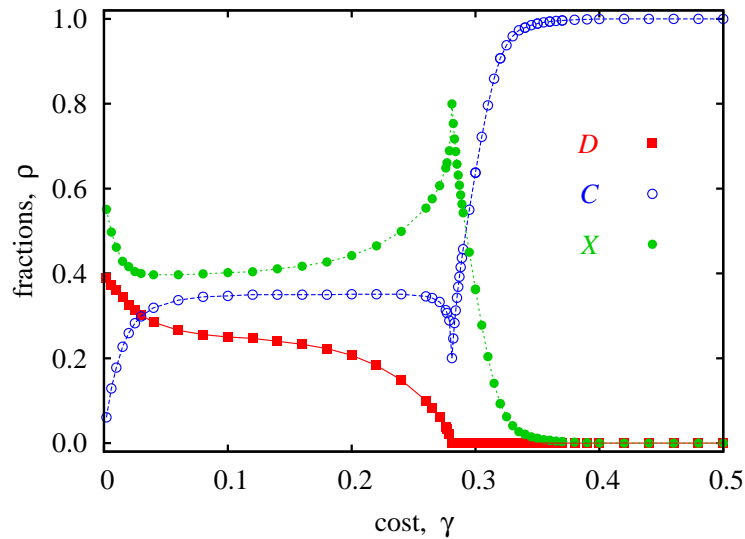,width=10.52cm}}
\caption{Cross-section of the phase diagram depicted in Fig.~\ref{phasea}, as obtained for $p=1$. Depicted are stationary fractions of the three competing strategies in dependence on the cost of deceit $\gamma$. As the value of $\gamma$ increases, the three-strategy $C+D+X$ phase first gives way to the two-strategy $C+X$ phase, and subsequently to the absorbing $C$ phase. In this cross-section all phase transitions are continuous. We emphasize that the rise of the fraction of $X$ as $\gamma$ increases (before the extinction of $D$) is an unexpected and counterintuitive evolutionary outcome that can only be explained by means of the spontaneous emergence of cyclic dominance amongst all three competing strategies, as illustrated in Fig.~\ref{snaps}.}
\label{cost}
\end{figure}

Representative cross-sections of the phase diagram provide a more quantitative insight into the different phase transitions depicted in Fig.~\ref{phasea}. In Fig.~\ref{cost}, we first show how the fractions of the three strategies vary in dependence on the cost $\gamma$ at $p=1$, where conditional cooperators are 100\% effective in identifying pure defectors. When the cost is small, all three strategies coexist in a stable $D+C+X$ phase. As $\gamma$ increases, deceitful defectors initially suffer, but the actual victims turn out to be the pure defectors --- the main rivals of the deceitful defectors. Based on the presented results, we may conclude that, up to a certain point, deceitful behavior fares better if it is costly. Put differently, the larger the value of $\gamma$, the higher the fraction of strategy $X$ in the stationary state. Only after $D$ die out does the trend reverse, and larger values of $\gamma$ actually have the expected impact of lowering the evolutionary success of deceitful defectors, to the point when the latter finally die out to give rise to the absorbing $C$ phase.

\begin{figure}
\centerline{\epsfig{file=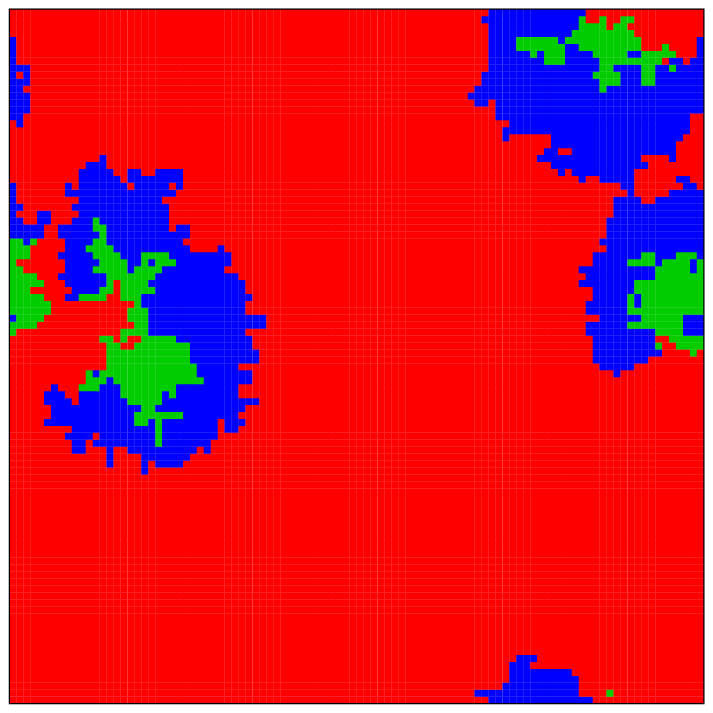,width=6cm}\epsfig{file=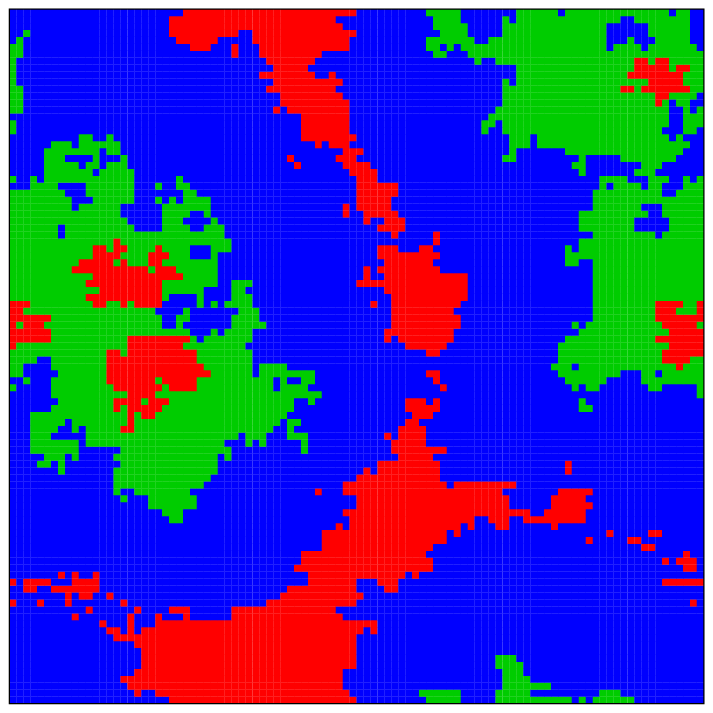,width=6cm}}
\centerline{\epsfig{file=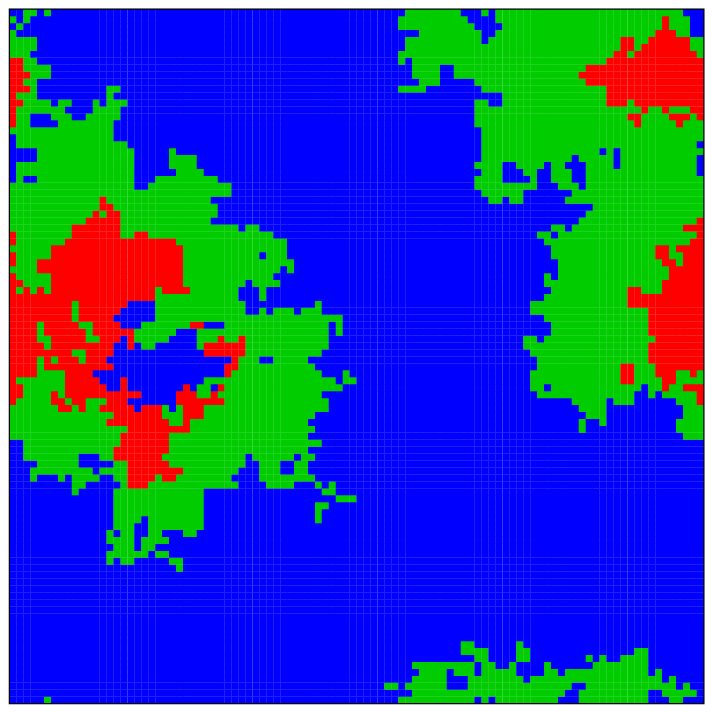,width=6cm}\epsfig{file=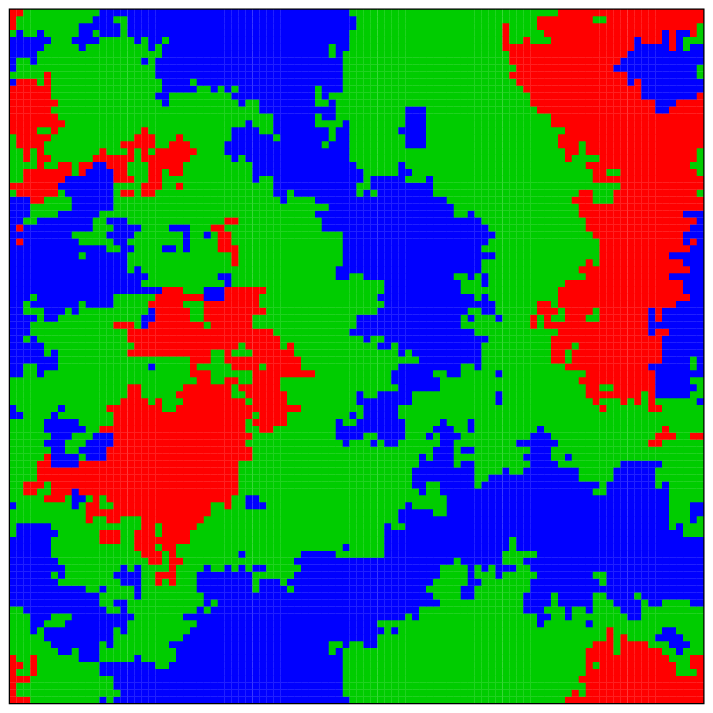,width=6cm}}
\caption{Consecutive snapshots of the square lattice, illustrating the spontaneous emergence of cyclic dominance from a random initial state between deceitful defectors (green), conditional cooperators (blue), and the pure defectors (red). The snapshots are taken at $60$, $100$, $120$ and $160$ MCS from top left to bottom right, respectively. Invasions proceed according to the $C \to D \to X \to C$ closed loop of dominance. Parameter values are: $r=0.3$, $p=1$, $\gamma=0.02$, and $L=100$.}
\label{snaps}
\end{figure}

This evolutionary paradox, namely that deceitful behavior fares better if it is costly, can only be explained through the self-organized spatial patterns that emerge spontaneously and drive cyclic dominance among the three competing strategies. As shown in Fig.~\ref{snaps}, traveling waves indeed emerge, where $C$ beats $D$, $D$ beats $X$, and $X$ beats $C$ to close the loop of dominance. The $C \to D \to X \to C$ loop of dominance is clearly inferable from the presented snapshots, as the initial $C$ wave (blue) spreads into the sea of $D$ (red). The pure defectors, on the other hand, invade the territory of $X$ (green), which in turn spread into the territory of $C$.

\begin{figure}
\centerline{\epsfig{file=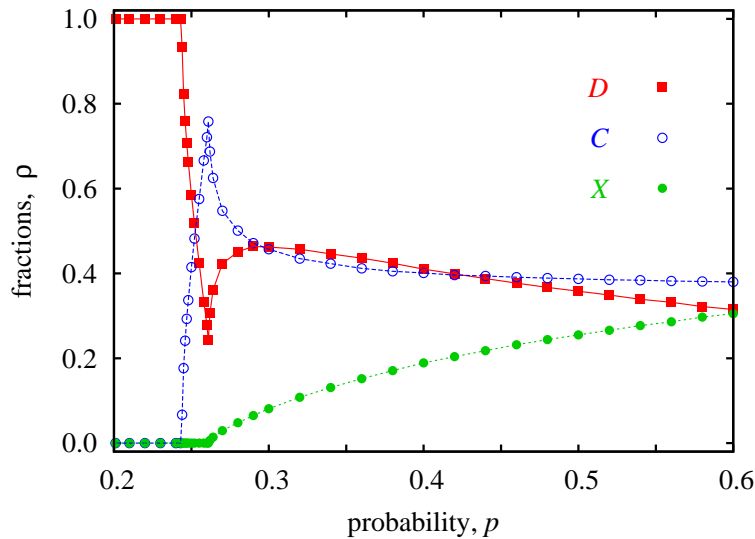,width=10.52cm}}
\caption{Cross-section of the phase diagram depicted in Fig.~\ref{phasea}, as obtained for $\gamma=0.2$. Depicted are stationary fractions of the three competing strategies in dependence on the probability $p$. As the value of $p$ increases, the absorbing $D$ phase first gives way to the two-strategy $C+D$ phase, and subsequently to the three-strategy $C+D+X$ phase. In this cross-section all phase transitions are continuous. We emphasize that the rise of the fraction of $D$ as $p$ increases (just after the emergence of $X$) is again an unexpected and counterintuitive evolutionary outcome that can only be explained by means of the spontaneous emergence of cyclic dominance amongst all three competing strategies (see main text for details).}
\label{prob1}
\end{figure}

\begin{figure}
\centerline{\epsfig{file=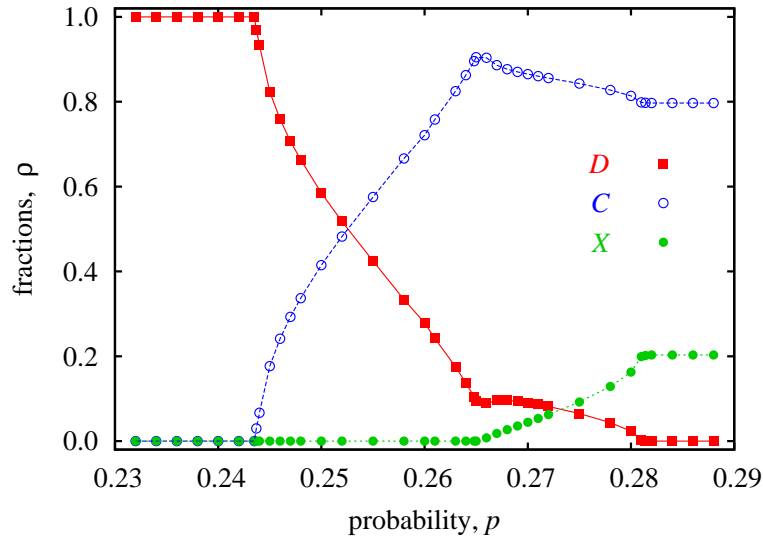,width=10.52cm}}
\caption{Cross-section of the phase diagram depicted in Fig.~\ref{phasea}, as obtained for $\gamma=0.31$. Depicted are stationary fractions of the three competing strategies in dependence on the probability $p$. As the value of $p$ increases, the absorbing $D$ phase first gives way to the two-strategy $C+D$ phase, then to the three-strategy $C+D+X$ phase, and finally to the two-strategy $C+X$ phase. Evidently, sufficiently increasing the value of $p$ may eradicate pure defectors and thus pave the way for deceitful defectors to capitalize on their investment $\gamma$. Due to network reciprocity, however, conditional cooperators never die out but rather form the $C+X$ phase.}
\label{prob2}
\end{figure}

Returning to the cross-sections of the phase diagram depicted in Fig.~\ref{phasea}, we show in Figs.~\ref{prob1} and \ref{prob2} how the fractions of the three strategies vary in dependence on the probability $p$ at $\gamma=0.2$ and $\gamma=0.31$, respectively. If the probability to reveal $D$ is small, then conditional cooperators are unable to survive. Consequently, deceitful defectors do not exist either, as their ``targets'' ($C$) are not available, and the direct competition with $D$ obviously leaves them at a disadvantage due to non-zero $\gamma$. As $p$ increases, the absorbing $D$ phase gives way to the two-strategy $C+D$ phase, which is possible due to network reciprocity and is thus a purely spatial effect. Interestingly, weakening $D$ further by elevating $p$ will initially generate more $D$ players in the stationary state. As the value of $p$ increases further $D$ do begin to decline on the expense of $C$, but on the other hand, $X$ emerges and serves as an additional target for $D$. One might expect that increasing $p$ further will support $C$ because they will not let $D$ players exploit them. While the fraction of $D$ indeed decreases, this in turn paves the way for deceitful defectors, who can finally capitalize on their investment $\gamma$. Together, these ``plus'' and ``minus'' effects will nullify each other and leave the fraction of conditional cooperators practically unaffected, despite of their elevated efficiency in detecting pure defectors. Thus, again unexpectedly, a higher success rate of identifying defectors does not necessarily favor cooperative behavior.

Notably, a qualitatively different final phase is reached if we apply a higher value of $\gamma$, as shown in Fig.~\ref{prob2}. Here the fraction of $X$ cannot raise as high, which in turn provides less targets for pure defectors, who therefore die out more easily. In the absence of $D$, however, the conditional cooperators and deceitful defectors can coexist, which is again made possible by the clustering of cooperators and is thus a purely spatial effect. Naturally, if we increase the cost further, then strategy $X$ cannot survive either, and the population evolves from an absorbing $D$ to the absorbing $C$ phase via a coexisting $C+D$ phase (cross-section not shown).

\begin{figure}
\centerline{\epsfig{file=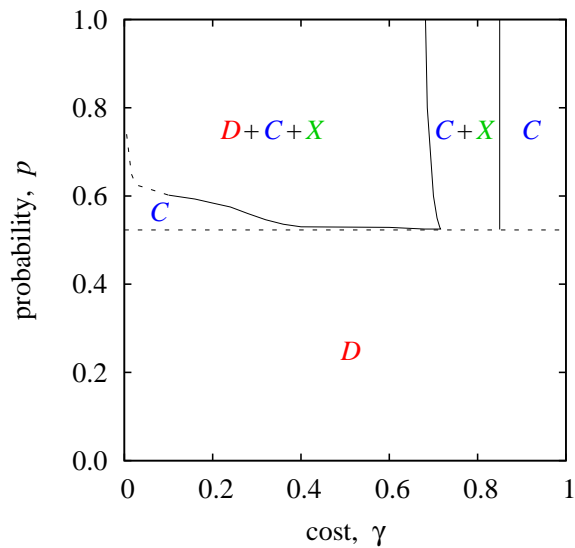,width=8cm}}
\caption{Full $\gamma-p$ phase diagram, as obtained for $r=0.7$. Solid lines denote continuous phase transitions, while dashed lines denote discontinuous phase transitions. As for the $r=0.3$ case depicted in Fig.~\ref{phasea}, here too the stable solutions include the three-strategy $C+D+X$ phase and the two-strategy $C+X$ phase, as well as the absorbing $D$ and $C$ phase. Evidently, there exist solutions that are independent of the strength of the social dilemma, but there also exist significant differences, like the nature of the phase transition points and the ``replacement'' of the two-strategy $C+D$ phase with the absorbing $C$ phase.}
\label{phaseb}
\end{figure}

In the strong limit of the prisoner's dilemma game that is obtained for $r=0.7$, we observe solutions that are qualitatively different from those obtained for $r=0.3$. Due to the large temptation to defect, pure defectors and conditional cooperators are unable to coexist in the absence of deceitful defectors. Instead, below $p_c=0.5231$, pure defectors will prevail, while above this critical value conditional cooperators will dominate the whole population. Remarkably, the consideration of deceitful defectors not only results in the lack of the $C+D$ phase, but gives rise also to the emergence of an absorbing $C$ phase at an intermediate value of $p$, even if the cost of deception is moderate. The phase diagram presented in Fig.~\ref{phaseb} summarizes these fascinating evolutionary outcomes, which indicate that belying cooperation may actually beget cooperation.

If we compare the two phase diagrams in Fig.~\ref{phasea} and Fig.~\ref{phaseb}, then we also find that there exist certain solutions that remain valid independently of the strength of the social dilemma. In addition to the dominance of $D$ at low values of $p$ and the dominance of $C$ at sufficiently high values of $\gamma$ and $p$, our previous observation regarding the optimal value of $\gamma$ when $C$ are efficient also remains valid. Namely, from the point of view of $X$ it is actually better to bear a larger cost of deceit than a small one, because the former effectively prevents pure defectors to exploit these additional efforts aimed at deceiving cooperators. In the absence of $D$, or when they are rare, the evolutionary advantage of $X$ can still manifest even at relatively large $\gamma$ values, especially if $r$ is also large. Moreover, for $r=0.7$ too, it is possible to observe that the highest detection probability does not always ensure the highest density of conditional cooperators. Even more strikingly, here an intermediate value of $p$ can result in an absorbing $C$ phase that becomes unstable at higher $p$ values.

\begin{figure}
\centerline{\epsfig{file=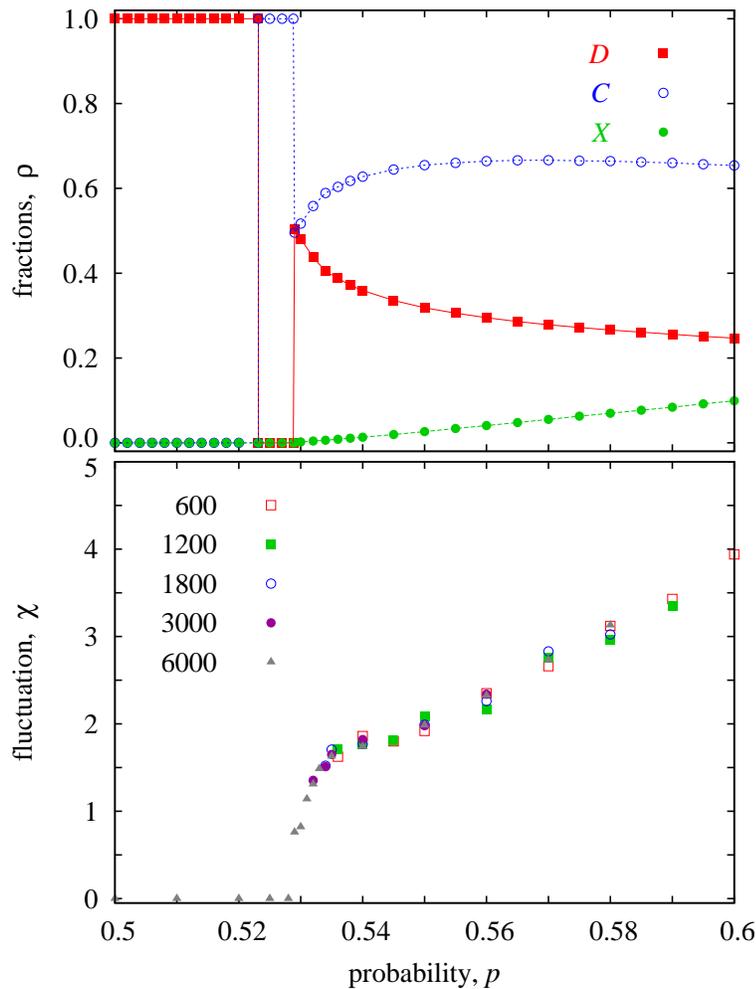,width=10.52cm}}
\caption{Top panel shows the cross-section of the phase diagram depicted in Fig.~\ref{phaseb}, as obtained for $\gamma=0.6$. Depicted are stationary fractions of the three competing strategies in dependence on the probability $p$. As the value of $p$ increases, the absorbing $D$ phase changes abruptly to the absorbing $C$, followed by the gradual emergence of deceitful defectors that give rise to the three-strategy $C+D+X$ phase. The bottom panel shows the fluctuation of the vanishing density of deceitful defectors $\rho_X$ (see Eq.~\ref{eq:chi}) in dependence on $p$, as obtained for different system sizes that are indicated in the figure legend. The always finite value of $\chi$ indicates that the amplitude of oscillations can always be reduced by increasing the system size, thus confirming a continuous phase transition.}
\label{pt1}
\end{figure}

\begin{figure}
\centerline{\epsfig{file=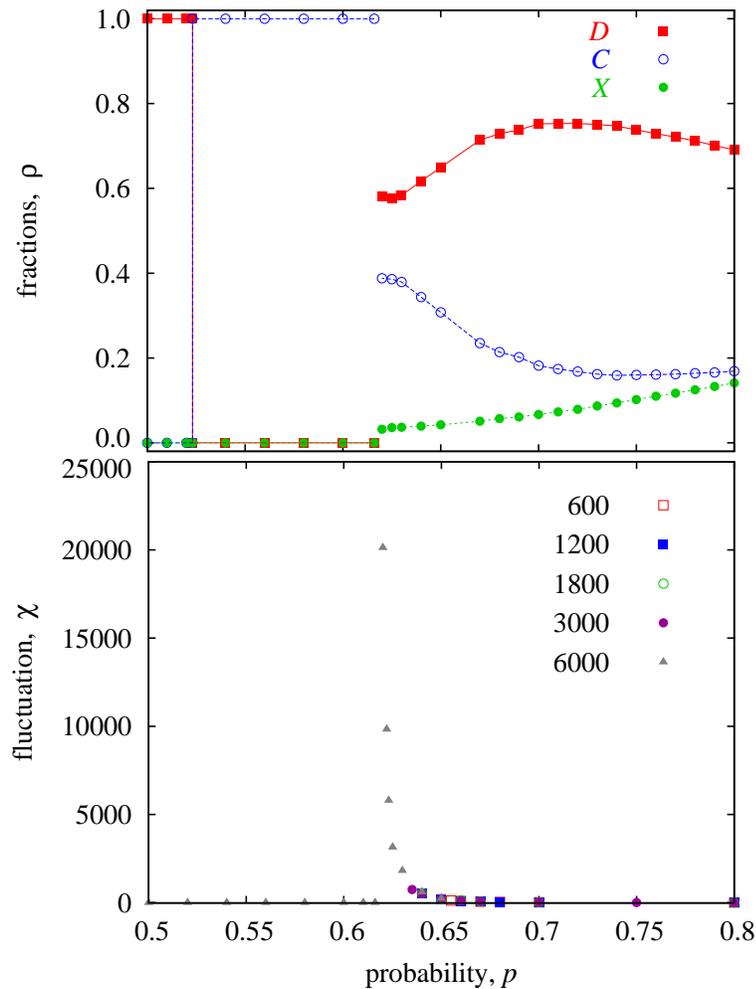,width=10.52cm}}
\caption{Top panel shows the cross-section of the phase diagram depicted in Fig.~\ref{phaseb}, as obtained for $\gamma=0.05$. Depicted are stationary fractions of the three competing strategies in dependence on the probability $p$. As the value of $p$ increases, the absorbing $D$ phase changes abruptly to the absorbing $C$, followed again by an abrupt emergence of deceitful defectors that give rise to the three-strategy $C+D+X$ phase. The bottom panel shows the fluctuation of the vanishing density of deceitful defectors $\rho_X$ in dependence on $p$, as obtained for different system sizes that are indicated in the figure legend. Unlike for $\gamma=0.6$ (see Fig.~\ref{pt1}), here the value of $\chi$ at the phase transition point is diverging, increasing beyond bound even for very large $L$, and thus indicating a fascinating discontinuous phase transition between the absorbing $C$ phase and the heterogeneous $C+D+X$ phase.}
\label{pt2}
\end{figure}

Also worthy of attention are the phase transitions between the three-strategy $C+D+X$ phase and the absorbing $C$ phase. When the cost of deception is large, then as $p$ decreases the frequency of $X$ decreases gradually, as shown in the top panel of Fig.~\ref{pt1}. When $X$ finally vanish, the competition between the remaining $D$ and $C$ terminates in an absorbing $C$ phase. As $p$ decreases further, an abrupt transition to an absorbing $D$ phase occurs. However, if the cost $\gamma$ is small, a qualitatively different behavior can be observed, as shown in the top panel of Fig.~\ref{pt2}. In this case, the average frequency of $X$ within the $C+D+X$ phase remains nonzero, but the amplitude of oscillations increases drastically as we approach the phase transition point.

Importantly, the increase in the amplitude of oscillations is not a finite size effect because the amplitude grows even if we increase the system size. This effect can be quantified by measuring the fluctuations of strategy $X$ according to
\begin{equation}
\chi= {L^2 \over M} \sum_{t_i = 1} ^M \left( \rho_X(t_i)- \overline{\rho_X} \right)^2 \,,
\label{eq:chi}
\end{equation}
where $M$ denotes the number of independent values measured in the stationary state. As the bottom panel of Fig.~\ref{pt1} demonstrates, this quantity remains finite at large $\gamma$, which means that the amplitude of oscillations can always be reduced by increasing the system size $L$. The same quantity, however, behaves very differently at small $\gamma$. As the bottom panel of Fig.~\ref{pt2} shows, the value of $\chi$ is diverging as we approach the phase transition point, indicating that here the amplitude of oscillations cannot be lowered and that $X$ will inevitably die out. We note that a similar type of discontinuous phase transition was already observed in the spatial public goods game with correlated positive and negative reciprocity, where cyclical dominance also emerged spontaneously between the competing strategies \cite{szolnoki_prx13}. These results thus reveal the hidden complexity behind the evolution of deception, which appears to be commonplace in evolutionary settings with three or more strategies in structured populations.

\section{Discussion}
To summarize, we have shown that the introduction of conditional cooperators and deceitful defectors to a social dilemma gives rise to many counterintuitive evolutionary outcomes that can only be understood as a consequence of self-organized pattern formation in structured populations. Spatial systems, where players have a limited interaction range, allow the observation of rich  behavior, including the formation of propagating fronts due to the spontaneous emergence of cyclic dominance. In particular, we have demonstrated the stable coexistence of all three competing strategies, as well as the emergence of $C+D$ and $C+X$ two-strategy phases. Similarly to the results obtained in well-mixed populations, in structured populations absorbing phases are also possible at specific parameter values. Namely, if the conditional cooperators are ineffective in identifying pure defectors, the latter dominate like under mean-field conditions. Conversely, if the conditional cooperators are sufficiently effective and if the cost of deception is high, then cooperative behavior dominates. Unexpectedly, our research indicates that an imperfect ability of cooperators to properly detect defectors may be beneficial for the evolution of cooperation, and that deceitful behavior may fare better if it is costly. These results are rooted in the spontaneous emergence of cycling dominance and complex spatial patterns. It is also worth emphasizing that these results are robust and remain valid if we choose other strategy updating rules or interaction networks other than lattices. For example, qualitatively similar evolutionary outcomes can be obtained on random regular graphs.

We have also shown that continuous and discontinuous phase transitions separate the different stable solutions, and that changing a model parameter may have highly nontrivial and unexpected consequences. For example, the three-strategy $C+D+X$ phase may terminate continuously or abruptly, depending solely on the cost of deception. This type of complexity is absent in traditional rock-scissors-paper or extended Lotka-Volterra-type models where the cyclic dominance is hardwired in the food web \cite{interface14}. Although the evolution of deception is also governed by cyclic dominance, the latter emerges spontaneously, and therefore changing even a single parameter may influence the effective invasion rates between all three competing strategies. We have made similar observations in the realm of the spatial ultimatum game \cite{szolnoki_prl12} and the spatial public goods game with reward \cite{szolnoki_epl10} and different types of punishment \cite{szolnoki_pre11b,sasaki_prsb13}, as well as in the spatial public goods game with correlated positive and negative reciprocity \cite{szolnoki_prx13}. Taken together, this indicates that the reported evolutionary complexity is in fact much more frequent than it might be assumed, and in fact appears to be commonplace in evolutionary settings where three or more strategies compete in a structured population.

When introducing punishing cooperators to a social dilemma, the tension between defectors and cooperators shifts to the tension between the cooperators that punish and the cooperators that do not punish. Since the latter avoid the additional costs that need to be invested for sanctioning defectors, they become second-order free-riders \cite{fehr_n04}. The evolution of deception gives rise to a conceptually similar shift in the dilemma, only that it is reversed and applies from the point of view of defectors. The main competitors of deceitful defectors are no longer cooperators, but rather the pure defectors who do not invest in deception. Whether it pays to defect or not depends on the effectiveness of the cooperators to correctly identify defectors. If conditional cooperation is effective, defectors can survive only if they hide behind those who actually make an effort in deceiving cooperators. Ultimately, this establishes the possibility of cyclic dominance between the three competing strategies, and it gives rise to the reported unexpected evolutionary outcomes. Specifically, for defectors it may be advantageous to bear a significant cost aimed at deceiving cooperators, although this depends nontrivially also on the efficiency of conditional cooperators.

Recent reviews attest to the fact that interdisciplinary approaches, linking together knowledge from biology, ecology, sociology as well as applied mathematics and physics, are successful in identifying new ways by means of which social dilemmas can be resolved in favor of cooperation \cite{nowak_s06,szabo_pr07,perc_bs10,perc_jrsi13}, and they also help to understand the observed complexity behind the riddles of evolution. We hope that our study will inspire further research aimed at investigating the role of deceitful strategies in evolutionary games, and we also hope that more experimental work will be carried out to clarify their role by the evolution of human cooperation \cite{d-orsogna_pone13,rand_tcs13,mitkidis_pone13,leibbrandt_pnas13}.

\ack
We thank our Referees for their insightful and constructive comments. This research was supported by the Hungarian National Research Fund (Grant K-101490), TAMOP-4.2.2.A-11/1/KONV-2012-0051, and the Slovenian Research Agency (Grant J1-4055).

\section*{References}
\providecommand{\newblock}{}


\begin{thebibliography}{10}
\expandafter\ifx\csname url\endcsname\relax
  \def\url#1{{\tt #1}}\fi
\expandafter\ifx\csname urlprefix\endcsname\relax\def\urlprefix{URL }\fi
\providecommand{\eprint}[2][]{\url{#2}}

\bibitem{carroll_06}
Carroll SB (2006) {\em The Making of the Fittest\/} (New York: W. W. Norton \& Company)

\bibitem{owen_80}
Owen D (1980) {\em Camouflage and Mimicry\/} (Chicago: The University of Chicago Press)

\bibitem{davies_00}
Davies NB (2000) {\em Cuckoos, Cowbirds and other Cheats\/} (Princeton: Princeton University Press)

\bibitem{macy_pnas02}
Macy MW, Flache A (2002) {\em Proc Natl Acad Sci USA\/} {\bf 99} 7229--7236

\bibitem{hardin_g_s68}
Hardin G (1968) {\em Science\/} {\bf 162} 1243--1248

\bibitem{axelrod_84}
Axelrod R (1984) {\em The Evolution of Cooperation\/} (New York: Basic Books)

\bibitem{nowak_11}
Nowak MA, Highfield R (2011) {\em SuperCooperators: Altruism, Evolution, and Why We Need Each Other to Succeed\/} (New York: Free Press)

\bibitem{nowak_s06}
Nowak MA (2006) {\em Science\/} {\bf 314} 1560--1563

\bibitem{weibull_95}
Weibull JW (1995) {\em Evolutionary Game Theory\/} (Cambridge, MA: MIT Press)

\bibitem{hofbauer_98}
Hofbauer J, Sigmund K (1998) {\em Evolutionary Games and Population Dynamics\/} (Cambridge, U.K.: Cambridge University Press)

\bibitem{nowak_06}
Nowak MA (2006) {\em Evolutionary Dynamics\/} (Cambridge, MA: Harvard University Press)

\bibitem{fudenberg_e86}
Fudenberg D, Maskin E (1986) {\em Econometrica\/} {\bf 54} 533--554

\bibitem{nowak_n93}
Nowak MA, Sigmund K (1993) {\em Nature\/} {\bf 364} 56--58

\bibitem{szabo_pr07}
Szab{\'o} G, F{\'a}th G (2007) {\em Phys Rep\/} {\bf 446} 97--216

\bibitem{santos_prl05}
Santos FC, Pacheco JM (2005) {\em Phys Rev Lett\/} {\bf 95} 098104

\bibitem{imhof_pnas05}
Imhof LA, Fudenberg D, Nowak MA (2005) {\em Proc Natl Acad Sci USA\/} {\bf 102} 10797--10800

\bibitem{tanimoto_pre07}
Tanimoto J (2007) {\em Phys Rev E\/} {\bf 76} 021126

\bibitem{gomez-gardenes_prl07}
G{\'o}mez-Garde{\~n}es J, Campillo M, Flor{\'{\i}}a LM, Moreno Y (2007) {\em Phys Rev Lett\/} {\bf 98} 108103

\bibitem{poncela_njp07}
Poncela J, G{\'o}mez-Garde{\~n}es J, Flor{\' \i}a LM, Moreno Y (2007) {\em New J Phys\/} {\bf 9} 184

\bibitem{fu_pre08b}
Fu F, Hauert C, Nowak MA, Wang L (2008) {\em Phys Rev E\/} {\bf 78} 026117

\bibitem{lee_s_prl11}
Lee S, Holme P, Wu ZX (2011) {\em Phys Rev Lett\/} {\bf 106} 028702

\bibitem{antonioni_pone11}
Antonioni A, Tomassini M (2011) {\em PLoS ONE\/} {\bf 6} e25555

\bibitem{fu_srep12}
Fu F, Tarnita C, Christakis N, Wang L, Rand D, Nowak MA (2012) {\em Sci Rep\/} {\bf 2} 460

\bibitem{tanimoto_pre12}
Tanimoto J, Brede M, Yamauchi A (2012) {\em Phys Rev E\/} {\bf 85} 032101

\bibitem{yang_njp14}
Yang H-X, Rong Z, Wang W-X (2014) {\em New J Phys\/} {\bf 16} 013010

\bibitem{szolnoki_epl11}
Szolnoki A, Xie NG, Wang C, Perc M (2011) {\em EPL\/} {\bf 96} 38002

\bibitem{brede_epl11}
Brede M (2011) {\em EPL\/} {\bf 94} 30003

\bibitem{fu_prsb11}
Fu F, Rosenbloom DI, Wang L, Nowak MA (2011) {\em Proc R Soc B\/} {\bf 278} 42--49

\bibitem{vukov_njp12}
Vukov J, Santos F, Pacheco J (2012) {\em New J Phys\/} {\bf 14} 063031

\bibitem{brede_acs12}
Brede M (2012) {\em Advs Complex Syst\/} {\bf 15} 1250074

\bibitem{szolnoki_pre12}
Szolnoki A, Perc M (2012) {\em Phys Rev E\/} {\bf 85} 026104

\bibitem{fu_srep12b}
Fu F, Nowak M, Christakis N, Fowler J (2012) {\em Sci Rep\/} {\bf 2} 845

\bibitem{sasaki_dga14}
Sasaki T (2014) {\em Dyn Games Appl\/} {\bf 4} 345--362

\bibitem{szolnoki_pre13}
Szolnoki A, Xie NG, Ye Y, Perc M (2013) {\em Phys Rev E\/} {\bf 87} 042805

\bibitem{exadaktylos_srep13}
Exadaktylos F, Esp{\'{\i}}n A, Bra{\~n}as-Garza P (2013) {\em Sci Rep\/} {\bf 3} 1213

\bibitem{trivers_11}
Trivers R (2011) {\em The Folly of Fools: The Logic of Deceit and Self-Deception in Human Life\/} (New York: Basic Books)

\bibitem{mcnally_prsb13}
McNally L, Jackson AL (2013) {\em Proc R Soc B} {\bf 280} 20130699

\bibitem{iniguez_prsb14}
I{\~n}iguez G, Govezensky T, Dunbar R, Kaski K, Barrio RA (2014) {\em Proc R Soc B} {\bf 281} 20141195

\bibitem{nowak_n92b}
Nowak MA, May RM (1992) {\em Nature\/} {\bf 359} 826--829

\bibitem{perc_bs10}
Perc M, Szolnoki A (2010) {\em BioSystems\/} {\bf 99} 109--125

\bibitem{santos_pnas06}
Santos FC, Pacheco JM, Lenaerts T (2006) {\em Proc Natl Acad Sci USA\/} {\bf 103} 3490--3494

\bibitem{santos_jtb12}
Santos FC, Pinheiro F, Lenaerts T, Pacheco JM (2012) {\em J Theor Biol\/} {\bf 299} 88--96

\bibitem{fortunato_pr10}
Fortunato S (2010) {\em Phys Rep\/} {\bf 486} 75--174

\bibitem{pinheiro_njp12}
Pinheiro F, Santos FC, Pacheco JM (2012) {\em New J Phys\/} {\bf 14} 073035

\bibitem{sun_njp10}
Sun J-T, Wang S-J, Huang Z-G, Yang L, Do Y, Wang Y-H (2010) {\em New J Phys\/} {\bf 12} 063034

\bibitem{szolnoki_pre11b}
Szolnoki A, Szab{\'o} G, Czak{\'o} L (2011) {\em Phys Rev E\/} {\bf 84} 046106

\bibitem{szolnoki_prl12}
Szolnoki A, Perc M, Szab{\'o} G (2012) {\em Phys Rev Lett\/} {\bf 109} 078701

\bibitem{szolnoki_prx13}
Szolnoki A, Perc M (2013) {\em Phys Rev X\/} {\bf 3} 041021

\bibitem{interface14}
Szolnoki A, Mobilia M, Jiang L-L, Szczesny B, Rucklidge AM, Perc M (2014) {\em J R Soc Interface\/} {\bf 11} 20140735

\bibitem{szolnoki_epl10}
Szolnoki A, Perc M (2010) {\em EPL\/} {\bf 92} 38003

\bibitem{sasaki_prsb13}
Sasaki T, Uchida S (2013) {\em Proc R Soc B\/} {\bf 280} 20122498

\bibitem{fehr_n04}
Fehr E (2004) {\em Nature\/} {\bf 432} 449--450

\bibitem{perc_jrsi13}
Perc M, G{\'o}mez-Garde{\~n}es J, Szolnoki A, Flor{\'{\i}a and Y Moreno} LM (2013) {\em J R Soc Interface\/} {\bf 10} 20120997

\bibitem{d-orsogna_pone13}
D'Orsogna M, Kendall R, McBride M, Short M (2013) {\em PLoS ONE\/} {\bf 8} e61458

\bibitem{rand_tcs13}
Rand DA, Nowak MA (2013) {\em Trends in Cognitive Sciences\/} {\bf 17} 413--425

\bibitem{mitkidis_pone13}
Mitkidis P, Sorensen J, Nielbo K, Andersen M, Lienard P (2013) {\em PLoS ONE\/} {\bf 8} e64776

\bibitem{leibbrandt_pnas13}
Leibbrandt A, Gneezy U, List JA (2013) {\em Proc Natl Acad Sci USA\/} {\bf 110} 9305--9308

\end{thebibliography}
\end{document}